\documentclass[journal]{IEEEtran}
\pdfoutput=1
\usepackage{amsmath}
\usepackage{amssymb}
\usepackage{graphicx}
\usepackage{epsfig}
\usepackage{subfigure}
\usepackage{mathrsfs}
\usepackage{fancyhdr}
\usepackage{color}
\usepackage{cite}
\usepackage[alwaysadjust]{paralist}
\allowdisplaybreaks

\usepackage{flushend}

\newtheorem{theorem}{Theorem}
\newtheorem{corollary}{Corollary}

\newtheorem{assumption}{Assumption}

\newcommand{\PP}{\mathbb{P}}
\newcommand{\E}{\mathbb{E}}
\newcommand{\pc}{p_c}

\newcommand{\sinr}{{\rm SINR}}
\newcommand{\snr}{{\rm SNR}}

\newcommand{\dd}{{\rm d}}

\newcommand{\stsets}[1]{\mathbb{#1}}
\newcommand{\R}{\stsets{R}}

\begin{document}
\graphicspath{{./Figures/}}

\title{Analytical Modeling of Uplink Cellular Networks}

\author{Thomas~D.~Novlan,~\IEEEmembership{Member,~IEEE,} 
        Harpreet~S.~Dhillon,~\IEEEmembership{Student Member,~IEEE,}
        Jeffrey~G.~Andrews,~\IEEEmembership{Fellow,~IEEE}
        \thanks{Manuscript received March 6, 2012; revised September 25, 2012, December 21,
2012; accepted February 12, 2013. The associate editor coordinating the review of this paper and approving it for publication was L. Sanguinetti.}
		\thanks{This research has been supported by AT\&T Laboratories and by NSF grant CIF-1016649. A part of this paper was presented at IEEE Globecom 2012~\cite{DhiNovC2012}.}
        \thanks{T. D. Novlan is with Samsung Telecommunications America, Richardson, Texas. (email: t.novlan@sta.samsung.com)}\thanks{H. S. Dhillon and J. G. Andrews are with the Department
of Electrical and Computer Engineering, The University of Texas at Austin, Austin,
TX, 78712 USA (e-mail: dhillon@utexas.edu, jandrews@ece.utexas.edu).}
}



\maketitle

\begin{abstract}
Cellular uplink analysis has typically been undertaken by either a simple approach that lumps all interference into a single deterministic or random parameter in a Wyner-type model, or via complex system level simulations that often do not provide insight into why various trends are observed.  This paper proposes a novel middle way using point processes that is both accurate and also results in easy-to-evaluate integral expressions based on the Laplace transform of the interference.  We assume mobiles and base stations are randomly placed in the network with each mobile pairing up to its closest base station. Compared to related recent work on downlink analysis, the proposed uplink model differs in two key features. First, dependence is considered between user and base station point processes to make sure each base station serves a single mobile in the given resource block. Second, per-mobile power control is included, which further couples the transmission of mobiles due to location-dependent channel inversion.
Nevertheless, we succeed in deriving the coverage (equivalently outage) probability of a typical link in the network. This model can be used to address a wide variety of system design questions in the future. In this paper we  focus on the implications for power control and see that partial channel inversion should be used at low signal-to-interference-plus-noise ratio ($\sinr$), while full power transmission is optimal at higher $\sinr$.
\end{abstract}


\section{\label{sec:intro} Introduction}
Modern cellular networks are evolving from voice-oriented to ubiquitous mobile-broadband data networks. While the downlink of these networks typically drives their bandwidth and speed requirements, improvements in uplink performance are increasingly important due to symmetric traffic applications like social networking, video-calls, and real-time generation and sharing of media content. A complete analytical framework for the cellular uplink requires several fundamental changes to the system model compared to the downlink, nearly all of which make analysis more difficult. While interference in the downlink comes from the fixed locations, in the uplink, interference is generated by mobile devices distributed throughout the network. A second distinction is the use of location dependent power control, which makes the transmit power highly variable, and therefore significantly changes the interference statistics compared to the downlink. Additionally, for the uplink, both a maximum power constraint and consideration of average transmit power are especially important for battery powered user devices. These constraints and their interdependence have made analysis of the uplink very challenging using traditional approaches. Using tools from stochastic geometry and point process theory, this work presents to our knowledge the first non-trivial tractable model for determining the fundamental metric of uplink performance, the complete $\sinr$ distribution, which immediately gives coverage and outage probability, and allows rate to be easily computed as well.

\subsection{Uplink Modeling Approaches}
One approach for analysis of the uplink has been to use the Wyner model \cite{Wyner94}. It is attractive for its analytical simplicity, wherein gains between users and base stations are normalized by the desired link and inter-cell interference is either a constant value or a single random variable to account for fading. Historically popular for evaluating the performance of code division multiple access (CDMA) based networks, this model is still used to evaluate performance from an information-theory perspective including multi-cell processing techniques \cite{Wyner2000,Sin2006,Simeone2009,Sanderovich2009,Levy2010,Onireti2011}. However, in \cite{Jiaming2011} the applicability and accuracy of the model is shown to be limited to scenarios where the interference can be spatially averaged, for example a CDMA network under high load. This type of interference averaging approach wherein inter-cell interference is assumed to be fixed is not a valid assumption for modern cellular systems where typically only a single user per cell (or sector) is active in a given resource block.

On the other extreme from the Wyner model, base stations are commonly modeled in a deterministic grid-based deployment, e.g., the popular hexagonal grid model. This approach does not lead to a tractable framework and results are based upon several simplifying approximations followed by exhaustive Monte Carlo simulations. Perhaps more importantly, the grid model -- which was always highly idealized -- is particularly out-of-touch with ongoing deployments, which have highly variable cell sizes and opportunistic placement of new towers \cite{Dohler2006,Qualcomm2011}. 

The inadequacy of existing approaches has led to an increased interest in the use of random spatial models for the network topology \cite{Haenggi2009}.  An advantage of this new approach is the ability to derive tractable expressions leading to more general performance characterizations and intuition \cite{Baccelli1995,Brown2000,AndrewsJ2011,MukherjeeWCDMA2011,MukherjeeLTE2011,Dhillon2011,Novlan2011,Novlan2012,HetNetCommMag2012}.  In our recent work \cite{AndrewsJ2011}, we showed that a completely random (Poisson) placement of base stations was about as accurate as a grid model, and sometimes more so, when compared to a large modern cellular network. Recent work in \cite{TaylorC2012} further illustrated that this model can provide more accurate $\sinr$ statistics for actual urban deployments. While this approach has mostly been applied to the downlink, recent work has also attempted to extend this to the uplink by deriving some approximate results for interference limited networks \cite{Govindasamy2011}. Under assumptions of multiple users per base station, transmitting on the same frequency in CDMA-style, and averaged transmit power for the interference, the authors of \cite{Govindasamy2011} derive analytical expressions for average spectral efficiency as the numbers of antennas, base stations, and users grow asymptotically large.

\subsection{Power Control}
Power control in various forms has been one of the key system design features for past, current, and proposed wireless standards \cite{Whitehead1993,Yates1995}. Fast uplink power control has been an especially important feature in CDMA-based networks \cite{Holma2001,Herdtner2000,Kim2001,Agrawal2005}. One reason for this is to mitigate the ``near-far'' problem that occurs when a base station cannot decode the signals of cell-edge users due to the much greater received power (and thus interference) caused by cell-interior users. For modern orthogonal frequency division multiple access (OFDMA) based cellular networks, due to the orthogonality of per-cell resources removing intra-cell interference and the aggressive use of adaptive modulation and coding techniques, fast power control is not as important of a feature. Instead, slow power control is typically considered, which attempts to overcome pathloss and large-scale fading (shadowing). For example, the 3GPP-LTE standard supports the utilization of open and closed-loop fractional power control in the uplink \cite{Ericsson2007,Ericsson2008}. While having an impact on coverage and rate, both by overcoming path loss and reshaping the distribution of the interference power, power control is also an important factor for battery utilization. Without a tractable analytical model it is difficult to gain intuition or derive quantitative results from a system/network design perspective due to the complicated relationship between the relevant system parameters. 


Recent work on the use of power control in modern OFDMA-based networks has focused on evaluating performance of different power control algorithms for a given set of system parameters via intensive simulations. The authors of \cite{Mullner2009} evaluated the impact of the maximum transmit power on open and closed-loop algorithms. In \cite{Castellanos2008}, the authors investigate the use of fractional power control in the uplink as a method for maintaining constant interference power at the base station. Maintaining constant interference power is of interest for many practical receiver algorithms which attempt to mitigate the impact of the interference but typically require either knowledge of the interference power or require it to be roughly constant. However, these studies, along with several earlier ones in \cite{Xiao2006,Rao2007}, utilize the standard regular hexagonal model for base station locations and the results are produced via simulation, which limits the scope to a limited set of possible design parameters.

Very recent work by the authors of \cite{Orange2011} proposes an analytical approach to this problem, with a particular goal of giving insight into the selection of fractional power control parameters. They consider a grid deployment for the base station locations and utilize a so-called ``fluid'' model which approximates the interference received in the center cell as coming from outside cells with the base stations located on rings of fixed radii \cite{Kelif2009}. Under these assumptions they derive expressions for the $\sinr$ and spectral efficiency for users located at relative cell-edge and cell-center locations, and use them to infer optimal power control parameters.

\subsection{Contributions}
The main contribution of this work is the derivation of uplink coverage probability for a randomly chosen mobile user with fractional power control, which is a general power control framework that incorporates virtually all modern cellular systems. We model the locations of the mobile users as a realization of the Poisson Point Process (PPP) and then assume that the base station corresponding to each mobile user is located uniformly in its Voronoi cell~\cite{Stoyan1996}. The uplink analysis is significantly more involved than its downlink counterpart because of this location dependence. Furthermore, the transmit power of a mobile in the uplink depends upon the distance to its associated base station due to the fractional power control. It turns out that the random variables denoting this distance for each mobile user are identically distributed but not independent in general. This dependence is not easy to model accurately and hence leads to further technical challenges in the derivation of coverage probability. However, we demonstrate that this dependence is weak and can be ignored, which improves the tractability of the system model with minimal impact on the accuracy of the results. Under this assumption, we derive analytical expression for the coverage probability of a randomly chosen mobile user. We further show that the same framework can be used to derive uplink coverage probability of a ``regular'' base station deployment by considering appropriate distribution of the distances of the interfering mobiles to their serving base stations.
Interestingly, this analytical result closely approximates the coverage probability computed numerically for the hexagonal grid model. We comment more on these observations in Section \ref{sec:cp}.


After a discussion of the derived expressions for coverage and average rate, we present system design guidelines comparing downlink and uplink coverage, and evaluate coverage probability and transmit power utilization as a function of the power control parameters. These results quantify the tradeoff between improved cell-edge $\sinr$ for low and moderate values of the fractional power control factor with significant overall power reduction available if power control is more aggressively applied. In the next section, we give our system model and discuss important underlying assumptions.

\begin{figure} [!ht]
\begin{center}
   \includegraphics[width=\columnwidth]{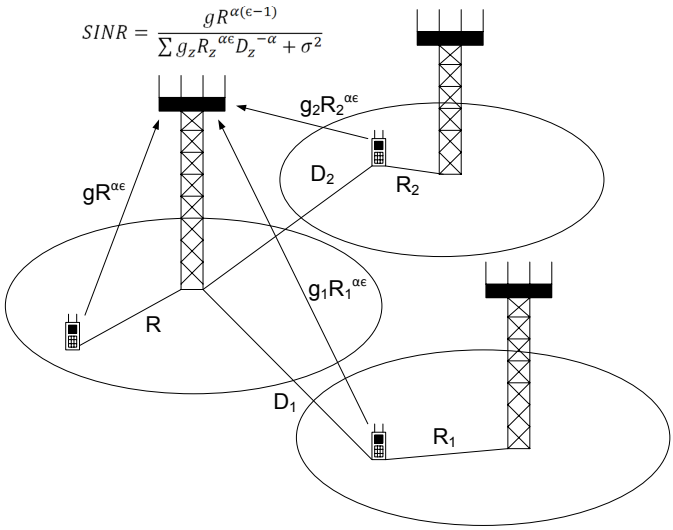}
   \caption{Visual system model example giving the $\sinr$ at the typical base station, focusing on the serving mobile and two interfering mobiles in adjacent cells.}
   \label{fig:sysMod}
   \end{center}
\end{figure}

\section{System Model \label{sec:sysmod}}
\subsection{System Setup and Modeling Assumptions}
We consider the uplink of a cellular network utilizing an orthogonal multiple access technique composed of a single class of base stations, macro base stations for example, and focus on the received $\sinr$ at a randomly chosen base station. Fig. \ref{fig:sysMod} gives a visual representation of the uplink system model and relationship between system parameters. The mobile user locations are assumed to form a realization of a homogeneous two-dimensional spatial PPP~\cite{Stoyan1996} with density $\lambda$. The spatial PPP corresponds to a uniform distribution of users in the network, which is the baseline assumption for many cellular system studies \cite{3GPP2009}. We assume that a mobile user is connected to the closest base station, corresponding to an association metric of max-received power averaged over fading, and that each base station has an active uplink user scheduled. Under such an assumption, it is reasonable to assume that each base station is uniformly distributed in the Voronoi cell of its corresponding mobile user. We perform analysis on a randomly chosen base station. The distance of this base station to the closest mobile is denoted by a random variable $R$. For tractability, we will approximate this choice with a case where the analysis is performed on a point uniformly chosen in $\mathbb{R}^2$. It is worth noting that there is a subtle difference between the random choice of a base station and a point uniformly chosen in $\mathbb{R}^2$ due to coupling induced by the dependence in the mobile and the base station point processes. To understand this difference, recall that a uniformly chosen point is biased to lie in bigger Voronoi cells more frequently, whereas there is no such bias when a base station is chosen randomly from the set of base stations defined as above. Nevertheless, we show that this base station selection approximation is tight in the context of the performance metrics of coverage and rate considered in this paper. The assumption is formally stated next.
\begin{assumption}
For tractability, we approximate the choice of a randomly chosen BS from a set of BSs defined such that exactly one BS falls in the Voronoi cell of each mobile by a point uniformly chosen in $\R^2$.
\end{assumption}

Under the above assumption, the random variable $R$ can be shown to be Rayleigh distributed.
The proof follows from the null probability of a two dimensional PPP \cite{Stoyan1996} as follows:
\begin{align}
\PP[R>r] &=\PP [\text{No mobile in circle of area $\pi r^2$}] = e^{-\lambda \pi r^2},
\end{align}
from which the probability density function (pdf) of $R$ follows:
\begin{equation}
f_{R}(r) = 2\pi\lambda r e^{-\lambda \pi r^2},\ r \geq 0. 
\label{eq:Rpdf}
\end{equation}

To model uplink interference, the randomly chosen base station is assumed to be located at the origin, which follows from the the translation invariance of the point process under Assumption 1. We denote the set of interfering mobiles by $\mathcal{Z}$, the distance of an interfering mobile $z \in \mathcal{Z}$ to the base station of interest by $D_z$, and the distance of the interfering mobile to its serving base station by $R_z$. It should be noted that the random variables $\{R_z\}_{z\in \mathcal{Z}}$ are identically distributed but not independent in general. The dependence is induced by the structure of Poisson-Voronoi tessellation and the restriction that only one base station can lie in each Voronoi cell. To visualize this dependence, recall a simple fact that the presence of a base station in a particular Voronoi cell forbids the presence of any other base station in that cell. However, as discussed in detail in the next section, this dependence is weak, which motivates the following independence assumption.
\begin{assumption}[Independence Assumption]
We assume that the random variables $\{R_z\}_{z\in \mathcal{Z}}$ are independent and identically distributed (i.i.d.).
\end{assumption}
For the marginal distribution of $R_z$, we make the following assumption, which lends tractability to uplink analysis both in the case of non-uniform and regular coverage regions. Both these cases will be discussed in detail in the next section.
\begin{assumption}[Distribution of $R_z$]
The random variable $R_z$ is modeled as:
\begin{enumerate}[\indent(a)]
\item Rayleigh distributed in the case when we study ``irregular'' deployment of base stations, i.e., non-uniform coverage regions, and,
\item distance of a point uniformly distributed over a circle of fixed radius from its center when we study regular deployment of the base stations, in particular the one corresponding to the hexagonal grid model.
\end{enumerate}
\end{assumption}
The Rayleigh distribution is motivated by the same null probability argument given in the case of $R$. Further details, e.g., the radius of the circle, and the motivation behind the second case will be given in the next section.

\subsection{Channel Model}
To model the channel, the path loss is assumed to be inversely proportional to distance with the path loss exponent given by $\alpha$. We consider small-scale Rayleigh fading between the mobiles and the base station under consideration, and a constant baseline mobile transmit power of $\mu^{-1}$. Thus the received power of the desired signal at the serving base station is given by $g R^{-\alpha\left(\epsilon - 1\right)}$, where $g$ is i.i.d. exponentially distributed with mean $\mu^{-1}$. The noise power is assumed to be ${\sigma}^2$. Next, we assume that all the mobiles utilize distance-proportional fractional power control of the form $R_z^{\alpha \epsilon}$, where $\epsilon \in [0, 1]$ is the power control factor. Thus, as a user moves closer to the desired base station, the transmit power required to maintain the same received signal power decreases, which is an important consideration for battery-powered mobile devices. Under this system model, the associated $\sinr$ at a base station located at origin is
\begin{equation}
\label{eq:SINR}
\sinr = \frac{g R^{\alpha\left(\epsilon - 1\right)}}{{\sigma}^2 + I_{\mathcal{Z}}},
\end{equation}
where for an interfering set of mobiles $\mathcal{Z}$,
\begin{equation}
\label{eq:IRUP}
I_{\mathcal{Z}} = \displaystyle\sum_{z\in \mathcal{Z}}{ \left(R_z^{\alpha}\right)^{\epsilon}g_z{D_z}^{-\alpha}}.
\end{equation}

If $\epsilon = 1$, the numerator of \eqref{eq:SINR} becomes $g$, with the pathloss completely inverted by the power control, and if $\epsilon = 0$ no channel inversion is performed and all the mobiles transmit with the same power. 


\subsection{Summary of Special Cases}
To validate the assumptions and highlight the importance of this model, we will compare the proposed model with various other approaches and special cases. For clarity, we describe all these approaches and special cases below:
\begin{asparadesc}
\item[PPP:] This corresponds to the proposed model without any assumptions, i.e., mobile locations correspond to a spatial PPP with a single base station dropped uniformly within the Voronoi cell of each mobile. Due to dependence induced by the structure of Poisson-Voronoi tessellation, direct analysis of this approach is daunting and hence not given. The numerical results will be provided for this case to validate various approximations leading to the following special cases.

\item[PPP-Rayleigh:] Setup is the same as the PPP case described above. For tractability we assume $R$ is Rayleigh distributed along with Assumptions $2$ and $3(a)$, i.e., $\{R_z\}$ are i.i.d. Rayleigh distributed. This case leads to the main result corresponding to the irregular base station deployment studied in  Section \ref{sec:PPPcase}.

\item[PPP-Uniform:] This case differs from PPP-Rayleigh only in terms of the distribution of $R_z$, which is as defined in assumption $3(b)$. In other words, we assume that the serving BS is located uniformly in a circle centered at the mobile user. This case leads to the main result corresponding to the regular base station deployment studied in Section \ref{sec:grid}.

\item[Grid:] To compare the results of the proposed tractable model, we consider the popular grid model, where the base stations are located on the centers of a hexagonal grid and one mobile user is distributed uniformly in each cell. Since this model does not lead to tractable expressions, it is evaluated via Monte-Carlo simulations.


\item[Log-normal:] This approach approximates inter-cell interference as a log-normal random variable with parameters determined through a numerical fit using simulations of the grid model. It has been utilized in previous works on coverage and capacity analysis for cellular networks due to favorable empirical evidence and its relative simplicity \cite{Tell99,Berg04}.
\end{asparadesc}

\begin{figure} [!ht]
\begin{center}
   \includegraphics[width=\columnwidth]{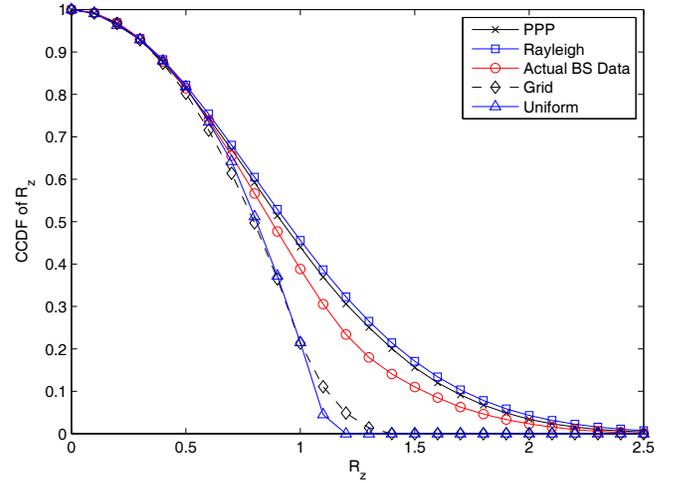}
   \caption{A comparison of the CCDFs of $R_z$ both for the PPP and a grid model with their respective approximations for $\lambda = 1/4$. Also included is the CCDF of $R_z$ for a set of real base station locations of an urban $4$G network.}
   \label{fig:RzComp}
   \end{center}
\end{figure}

\section{Coverage probability \label{sec:cp}}
The probability of coverage can be formally defined as the complementary cumulative distribution function (ccdf) of $\sinr$ as:
\begin{equation}
\pc = \PP[\sinr>T],
\end{equation}
which is the probability that the uplink $\sinr$ at a randomly chosen base station is greater than the target $\sinr$ $T$. It can also be visualized as being the average area or the average fraction of users in coverage. As noted earlier, we perform analysis on a randomly chosen base station assumed to be located at the origin that connects to the closest mobile user. Under assumption 1, the distribution of the distance of the closest mobile from the randomly chosen base station $R$ can be approximated by the Rayleigh distribution given by \eqref{eq:Rpdf}. As we discuss in detail for $R_z$ later in this section, this approximation is tight and does not affect the accuracy of our results.

The net interference at a randomly chosen base station is the sum of the powers from all the transmitting mobiles lying farther than $R$. Under the power control model described in the previous section, this power depends upon the distance of a mobile to its corresponding base station and the power control factor $\epsilon \in [0,1]$. For a mobile $z \in \mathcal{Z}$, we denote its distance to the corresponding base station as $R_z$. Although the random variables $\{R_z\}_{z\in \mathcal{Z}}$ are identically distributed, they are not independent in general. However, as shown later in this section, this dependence is weak and we will henceforth assume each $R_{z}$ to be i.i.d. (Assumption $2$).
Under this independence assumption, we first derive the coverage probability for the general distribution of $R_z$ and then use this general result to study two particular scenarios corresponding to non-uniform and regular coverage regions. The main uplink coverage probability result of this paper is stated in Theorem~\ref{thm:uplink1tier}. 


\begin{theorem}[Uplink coverage for i.i.d. $R_z$]
\label{thm:uplink1tier}
The uplink coverage probability is given by:
\begin{align}
& p_c(T,\lambda,\alpha,\epsilon) \nonumber \\
& = 2\pi\lambda \int_0^\infty r e^{-\pi\lambda r^2-\mu Tr^{\alpha(1-\epsilon)}\sigma^2}\mathscr{L}_{I_z}\left(\mu Tr^{\alpha(1-\epsilon)}\right) \dd r,
\label{eq:ulMain}
\end{align}
where the Laplace transform of the interference is given by $\mathscr{L}_{I_z}(s) =$
\begin{equation}
\label{eq:laplace}
\exp\left(-2\pi\lambda\int_{r}^{\infty}\left(1 - \E_{R_z}\left[\frac{\mu}{\mu + s R_z^{\alpha \epsilon} x^{-\alpha}}\right] \right)x \dd x\right).
\end{equation}
\end{theorem}

\begin{IEEEproof}
Starting from the definition of $p_c$ and $\sinr$, $p_c(T,\lambda,\alpha,\epsilon)$
\begin{align}
\label{eq:maineq}
&= \int_0^\infty \PP\left(\sinr > T\right)f_R(r) \dd r\\
&= \int_0^\infty \PP\left(\frac{g \left(r^{\alpha(\epsilon-1)}\right)}{{\sigma}^2 + I_{\mathcal{Z}}} > T\right)2\pi\lambda r e^{-\pi\lambda r^2} \dd r\\
&= \int_0^\infty \PP\left( g> \frac{T({\sigma}^2 + I_{\mathcal{Z}})}{r^{\alpha(\epsilon-1)} }\right)2\pi\lambda r e^{-\pi\lambda r^2} \dd r
\end{align}
\begin{align}
&\stackrel{(a)}{=}  \int_0^\infty 2\pi\lambda r e^{-\pi\lambda r^2}e^{-\mu Tr^{\alpha(1-\epsilon)}\sigma^2}\E_{I_z}\left[e^{-\mu Tr^{\alpha(1-\epsilon)}I_z}\right] \dd r\\
&\stackrel{(b)}{=}  \int_0^\infty 2\pi\lambda r e^{-\pi\lambda r^2}e^{-\mu Tr^{\alpha(1-\epsilon)}\sigma^2}\mathscr{L}_{I_z}\left(\mu Tr^{\alpha(1-\epsilon)}\right) \dd r,
\end{align}
where $(a)$ follows from the fact that $g \sim \exp(\mu)$ and $(b)$ follows from the definition of Laplace transform of interference $\mathscr{L}_{I_z}(s) = \E_{I_z}[e^{-sI_z}]$. To complete the proof, we now derive an expression for $\mathscr{L}_{I_z}(s)$ below:
\begin{align}
&\mathscr{L}_{I_z}(s) \nonumber \\
&= \E_{I_z}\left[\exp\left(-\sum_{z \in \mathcal{Z}}sR_z^{\alpha \epsilon}g_zD_z^{-\alpha}\right)\right]\\
&= \E_{R_z,g_z,D_z}\left[\prod_{z \in \mathcal{Z}}\exp\left(-s R_z^{\alpha \epsilon} g_zD_z^{-\alpha}\right)\right]\\
&\stackrel{(a)}{=} \E_{R_z,D_z}\left[\prod_{z \in \mathcal{Z}}\E_{g_z}\left[\exp\left(-s R_z^{\alpha \epsilon} g_zD_z^{-\alpha}\right)\right]\right]\\
&\stackrel{(b)}{=} \E_{D_z}\left[\prod_{z \in \mathcal{Z}}\E_{R_z}\left[\frac{\mu}{\mu + s R_z^{\alpha \epsilon} D_z^{-\alpha}}\right]\right] \\
&\stackrel{(c)}{=} \exp\left(-2\pi\lambda\int_{r}^{\infty}\left(1 - \E_{R_z}\left[\frac{\mu}{\mu + s R_z^{\alpha \epsilon} x^{-\alpha}}\right] \right)x \dd x\right),
\label{eq:laplace2}
\end{align}
where $(a)$ follows from the independence of $g_z$, $(b)$ follows from the independence of $R_z$ and from the fact that $g_z \sim \exp(\mu)$, and $(c)$ follows from the Probability Generating Functional (PGFL) of a PPP~\cite{Stoyan1996}.
\end{IEEEproof}

The coverage probability expression can be simplified for the full power control case ($\epsilon=1$) in the interference-limited scenario (inter-mobile interference dominates thermal noise), which is stated as the following corollary of Theorem \ref{thm:uplink1tier}.
\begin{corollary}
The uplink coverage probability for the full power control case ($\epsilon = 1$) assuming no noise ($\sigma^2=0$) is given by
\begin{equation}
p_c(T,\lambda,\alpha,\epsilon=1) = \int_0^\infty 2\pi\lambda r e^{-\pi\lambda r^2}\mathscr{L}_{I_z}\left(T\right) \dd r,
\end{equation}
where $\mathscr{L}_{I_z}(s)$ is a function of $r$ and is given by \eqref{eq:laplace2} with $\epsilon=1$ and $\mu=1$, where $\mu=1$ is due to the fact that when noise power is negligible, the $\sinr$ distribution is no longer a function of $\mu$.
\end{corollary}

\begin{figure} [!ht]
\begin{center}
   \includegraphics[width=\columnwidth]{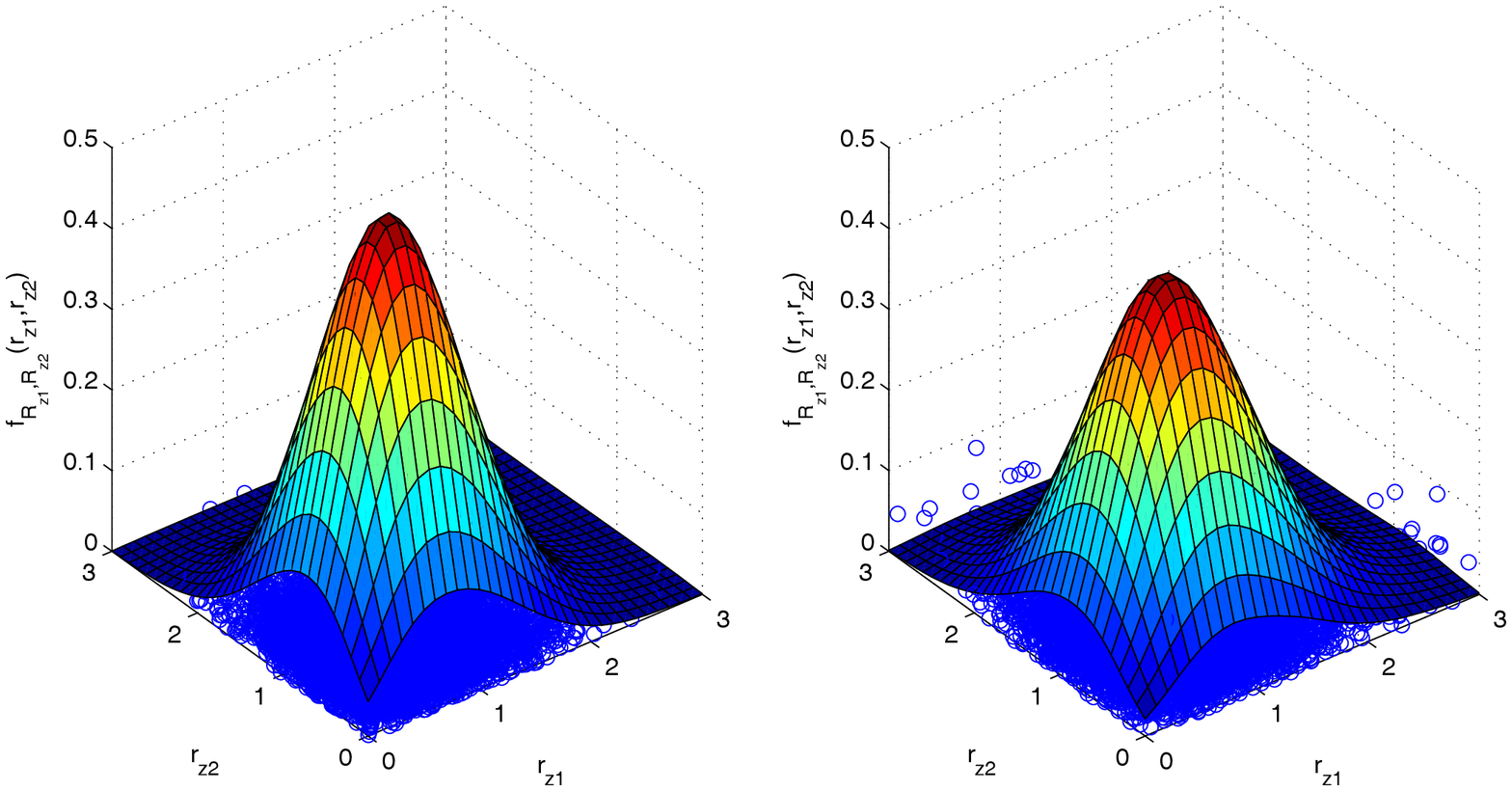}
   \caption{Joint densities of $R_{z_1}$ and $R_{z_2}$ for the actual PPP model (left) and under the independence assumption (right). $R_{z_1}$ and $R_{z_2}$ are the distances of the mobiles to their respective base stations in two neighboring Voronoi cells.}
   \label{fig:rzDensity3D}
   \end{center}
\end{figure}

\begin{figure} [!ht]
\begin{center}
   \includegraphics[width=\columnwidth]{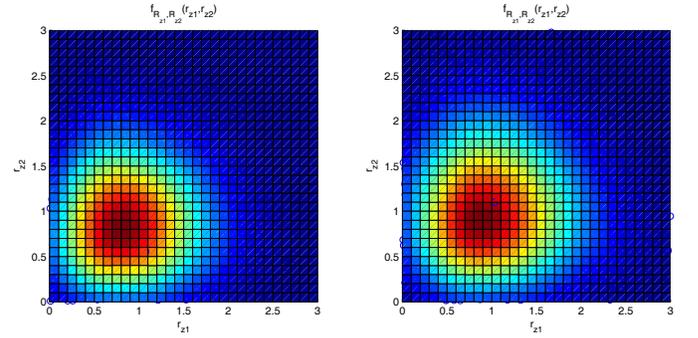}
   \caption{Top view of the joint densities of $R_{z_1}$ and $R_{z_2}$ for the actual PPP model (left) and under the independence assumption (right). $R_{z_1}$ and $R_{z_2}$ are the distances of the mobiles to their respective base stations in two neighboring Voronoi cells.}
   \label{fig:rzDensity2D}
   \end{center}
\end{figure}

\subsection{Distribution of $R_z$ and Comments on Independence Assumption}
\label{sec:PPPcase}
After deriving the coverage probability expressions for general $R_z$, we now derive the distribution of $R_z$ for the PPP model under the independence assumption. As mentioned in the previous section, each base station is randomly located in the Voronoi cell of its corresponding mobile. Therefore, as was done in case of $R$, $R_z$ can also be approximated by the distance of a randomly chosen point in $\mathbb{R}^2$ to its closest base station and hence its distribution can be approximated by Rayleigh distribution (as derived for $R$):
\begin{equation}
f_{R_z}(r_z) =  2\pi\lambda r_z e^{-\lambda \pi r_z^2}\ ,\ r_z \geq 0
\label{eq:pdfRz}
\end{equation}
The CCDF of $R_z$ is then $\PP[R_z > r_z] = e^{-\lambda \pi r_z^2}$, which is shown to be a tight fit for the numerical estimate for the PPP model in Fig.~\ref{fig:RzComp}. This corresponds to the Assumption $3(a)$ mentioned in the previous section and will be used to specialize Theorem~\ref{thm:uplink1tier} for the case of non-uniform coverage regions, i.e., irregular base station deployment. Although Fig.~\ref{fig:RzComp} shows that our approximations for the distributions of $R$ and $R_z$ are tight, it does not provide any insight into the extent of dependence between random variables $\{R_z\}_{z\in \mathcal{Z}}$ which is defined by their joint distribution. Since it is hard to gain insights from the complete joint distribution of $\{R_z\}_{z\in \mathcal{Z}}$, we study a simplified case of the joint distribution of two random variables $R_{z_1}$ and $R_{z_2}$, which are the distances of the mobiles to their respective base stations in two neighboring Voronoi cells. Since the dependence is expected to be strongest for the neighboring cells, this study can be thought of as a worst case study. We numerically compute the joint pdf $f_{R_{z_1}, R_{z_2}}(r_{z_1}, r_{z_2})$ for the actual PPP model and compare it with the joint pdf derived under the independence assumption in Fig.~\ref{fig:rzDensity3D}. It should be noted that the joint pdf under the independence condition follows directly from \eqref{eq:pdfRz} and is given by: $f_{R_{z_1}, R_{z_2}}(r_{z_1}, r_{z_2}) =$
\begin{equation}
(2\pi\lambda)^2 r_{z_1} r_{z_2} e^{-\lambda \pi (r_{z_1}^2 + r_{z_2}^2)}, r_{z_1} \geq 0,  r_{z_2} \geq 0.
\label{eq:pdfJRz}
\end{equation}
From Fig.~\ref{fig:rzDensity3D}, we note that the two joint densities are surprisingly similar, with the pdf slightly more dispersed in the case of the independence assumption, which is the expected direct result of independence. For better visualization, we also provide the top view of the joint densities in Fig.~\ref{fig:rzDensity2D}, which leads to the same conclusion. The correlation coefficient $\rho_{R_{z_1}, R_{z_2}}$ is numerically computed to be $0.07$ for this simulation setup.

\begin{figure}
\centering
\subfigure[$\alpha = 4,~\epsilon = 1$]{
\includegraphics[width=\columnwidth]{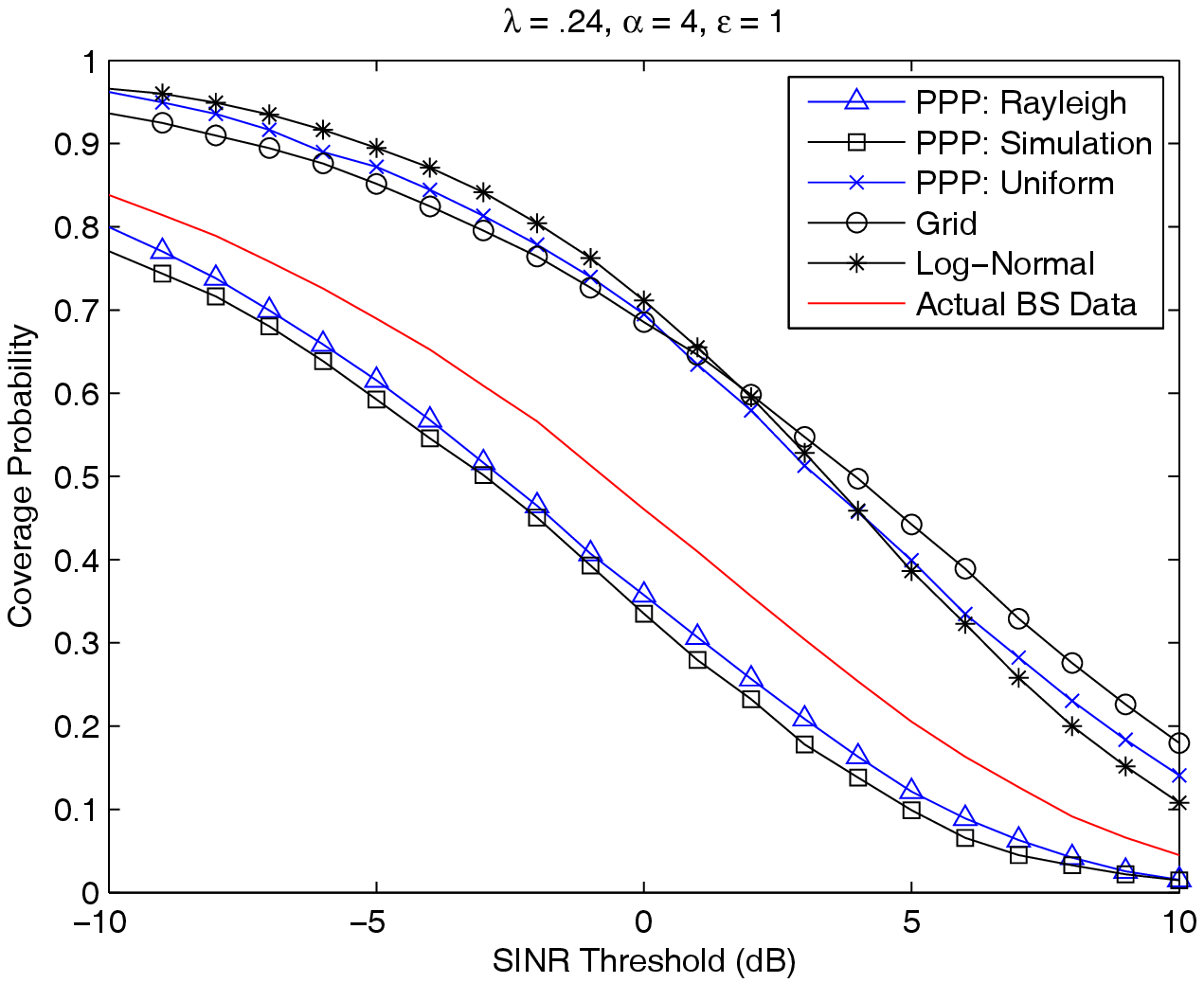}
\label{fig:Pcfull1}
}
\subfigure[$\alpha = 3.25,~\epsilon = .75$]{
\includegraphics[width=\columnwidth]{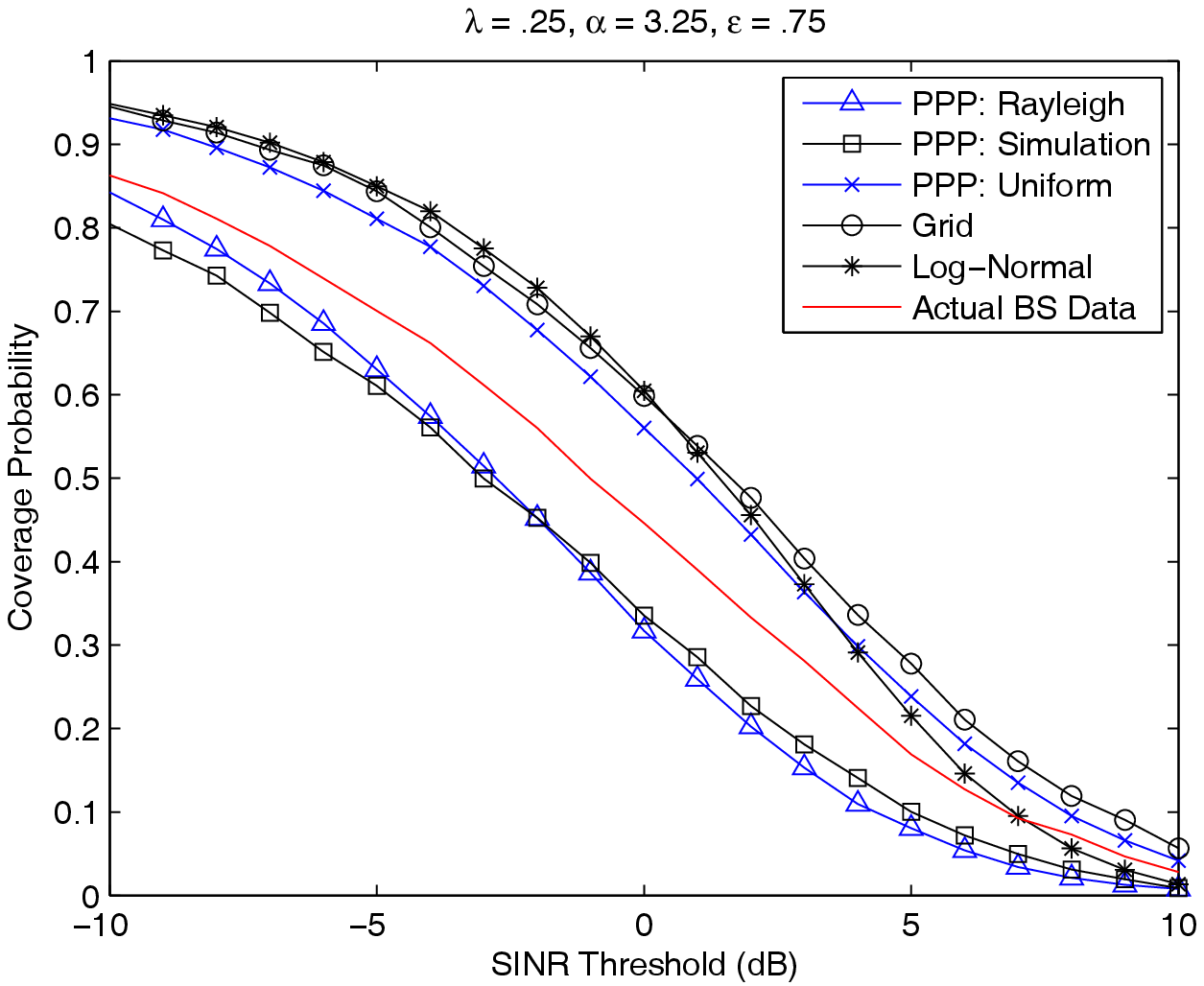}
\label{fig:Pcfull2}
}
\label{fig:Pcfull}
\caption{A comparison of the uplink coverage probability for the proposed model  with simulations for the grid and the PPP models. Also included is the result using a set of actual base station locations and results using a log-normal approximation for the interference.}
\end{figure}

After validating the independence assumption, we now use the density of $R_z$, given by \eqref{eq:pdfRz}, to derive the Laplace transform of interference for the PPP case, which is given by:
\begin{align}
&\mathscr{L}_{I_z}(s) =  \nonumber \\
&\exp\left(-2\pi\lambda\int\limits_{r}^{\infty}\left(1 - \int\limits_{0}^{\infty} \frac{\mu}{\mu + s u^\frac{\alpha \epsilon}{2} x^{-\alpha}} \pi \lambda e^{-\lambda \pi u} \dd u \right)x \dd x\right).
\label{eq:laplacePPP}
\end{align}
We plot the uplink coverage probability using this expression of the Laplace transform and compare it with the numerically computed coverage probability for a simulated PPP under true power control (without independence assumption) in Fig.~\ref{fig:Pcfull1} and Fig.~\ref{fig:Pcfull2} with no noise and $\lambda = .25$ for $\alpha = 4$, $\epsilon = 1$ and $\alpha = 3.25$, $\epsilon = .75$ respectively. We note that the analytical result derived under the independence assumption closely approximates the true power control result for a PPP as well as the results based on simulations utilizing a set of actual base station locations compared to simulations using the grid model.

The results in Fig.~\ref{fig:Pcfull1} and Fig.~\ref{fig:Pcfull2} are also further compared with a model which approximates inter-cell interference as a log-normal random variable with parameters determined through a numerical fit of the grid model. The log-normal approximation does not capture the shape of the $\sinr$ distribution as accurately as the proposed model. Approaches that combine the interference into a single term that must be empirically estimated cannot be easily parameterized as a function of key network features such as pathloss exponent, base station/user density, or fractional power control. However, since the proposed model is a function of these system parameters, we show in Sec. \ref{sec:sd} that it can be used to give insights into system design and performance trends.

\subsection{Comments on Regular (Grid) Model \label{sec:grid}}
Grid models are used to model more ``regular'' base station locations. The most popular model used in prior work places the base stations on a hexagonal grid. While this model has been extremely helpful in the numerical studies of macro-cellular networks, it does not provide analytical tractability. In this subsection, we show that the random spatial model for the mobile user locations along with an appropriately chosen distribution of $R_z$ provides a generative model that enables us to derive analytical expression for the coverage probability that closely approximates the numerically computed results for the grid model.

Approximating hexagons as circles with the same area $\lambda^{-1}$, we assume that each base station is located uniformly in a circle of radius $\frac{1}{\sqrt{\pi\lambda}}$ around its corresponding mobile. The radius value is evaluated from the density of the mobile users assuming there is one base station per mobile user. It is important to note that the only difference between this and the PPP-Rayleigh model studied in the previous subsection is the distribution of $R_z$. The difference is formally stated in Assumption $3$ in the previous section. The density of $R_z$ for this case can be easily evaluated as:
\begin{equation}
f_{R_z}(r_z) = 2 \pi \lambda r_z, r_z \in \left[0, \frac{1}{\sqrt{\pi\lambda}}\right].
\end{equation}
As shown in Fig. \ref{fig:RzComp}, this closely approximates the distribution of $R_z$ in a grid model. Using this density of $R_z$, we can now compute the Laplace transform of interference which can be expressed as: $\mathscr{L}_{I_z}(s) = $
\begin{equation}
\label{eq:laplaceUniform}
\exp\left(-2\pi\lambda\int_{r}^{\infty}\left(1 - \int_{0}^{\frac{1}{\sqrt{\pi \lambda}}} \frac{\mu}{\mu + s u^{\alpha \epsilon} x^{-\alpha}} 2 \pi \lambda u \dd u \right)x \dd x\right).
\end{equation}
While our results hold for general pathloss exponents $\alpha$, power control exponents $\epsilon$, and different noise powers $\sigma^2$, in the case of $\alpha = 4$ $\mu = 1$, and $\epsilon = 1$, the expression for the Laplace transform can be found in closed-form. In this case, closely corresponding to an interference-limited urban cellular deployment scenario \cite{Ericsson2008}, the Laplace transform is given as 
\begin{align}
\label{eq:laplaceUniform2}
\mathscr{L}_{I_z}(T) &= \exp\left(-\frac{\pi\lambda}{2}r^2 + \left(\frac{{\pi}^2{\lambda}^2\arctan\left(\frac{\sqrt{T}}{\pi\lambda r^2}\right)}{2\sqrt{T}}\right)r^4\right.\nonumber\\
&-\left. \frac{\sqrt{T}}{2}\arctan\left(\frac{\pi\lambda r^2}{\sqrt{T}}\right)\right).
\end{align}

We compare the coverage probability derived using this Laplace transform with the numerically computed coverage probability using true power control in a grid model in Fig.~\ref{fig:Pcfull1} and Fig.~\ref{fig:Pcfull2} with $\lambda = .25$ and $\sigma^2 = 0$ for $\alpha = 4$, $\epsilon = 1$ and $\alpha = 3.25$, $\epsilon = .75$. Surprisingly, we note that the analytical result derived using a random spatial model under the independence assumption in both cases closely approximates the true power control result for a hexagonal grid model even compared to the log-normal interference approximation. As with the PPP model, a crucial step is to appropriately choose the distribution of $R_z$. Thus, while utilizing the same underlying random spatial model for the mobile user locations, we are able to ``tune'' the results to fit a range of highly non-uniform to very regular network topologies.

\begin{figure} [!ht]
\begin{center}
   \includegraphics[width=\columnwidth]{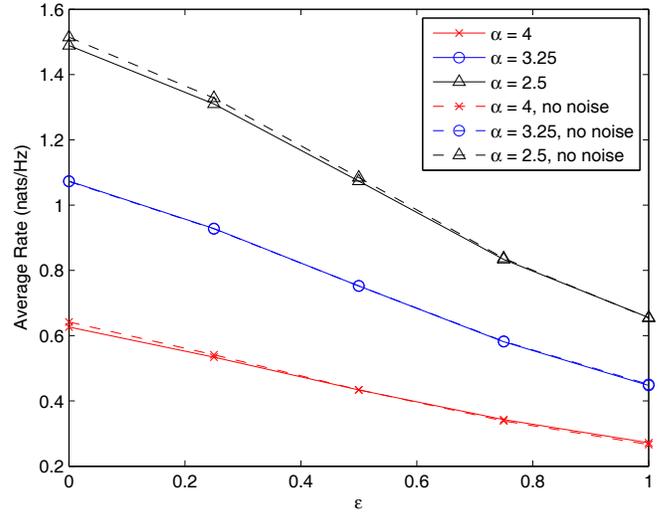}
   \caption{Average user rate as a function fractional power control parameter $\epsilon$ for pathloss exponents $\alpha = 2.5, 3.25$, and $4$, $\lambda = .24$, $\mu^{-1} = 200$ mW, and with no noise or with $\sigma^2 = -104$dBm.}
   \label{fig:ulRate}
   \end{center}
\end{figure}

\begin{figure} [!ht]
\begin{center}
   \includegraphics[width=\columnwidth]{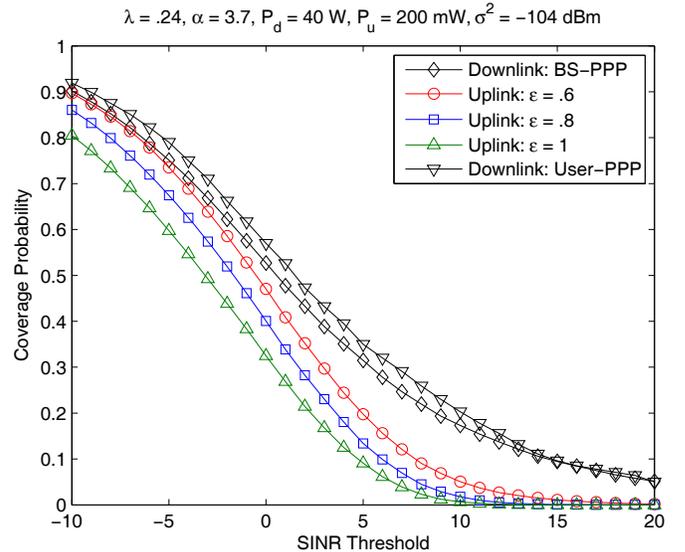}
   \caption{A comparison of the coverage probability for the downlink with 40W transmit power and the uplink utilizing fractional power control with $\epsilon = .6, .8$ and $1$, and a max transmit power of 200 mW.}
   \label{fig:DvU}
   \end{center}
\end{figure}

\section{System Design Applications \label{sec:sd}}
Based on the framework developed in Sec. \ref{sec:cp}, we analyze performance metrics in the context of realistic parameters for modern networks and gain insight into the system design. Here we primarily focus on the scenario of Sec. \ref{sec:PPPcase} where the base station is located uniformly in the Voronoi cell of its corresponding mobile user and the distance to the nearest base station is Rayleigh distributed. This is useful in capturing the non-uniform topology of many modern network deployments, although similar analysis could be performed for networks with regular topology as discussed in detail in the previous section.

\subsection{Average rate}
The use of link-adaptive algorithms in modern cellular networks allows the average $\sinr$ to be directly related to average data rate for mobile users. A straightforward application of the results of Sec. \ref{sec:cp} is to determine analytical expressions for user rate under different stochastic power control models as a function of the key uplink parameters, something previously not possible with deterministic network topology models.

Assuming adaptive modulation and coding, we define the average data rate based upon the Shannon capacity expression, $\bar{\tau}\left(\lambda,\alpha,\epsilon\right) = \mathbb{E}\left[\ln\left(1 + \sinr\right)\right]$, integrating over the $\sinr$ and fading distributions. For the sake of convenience, we give the results in units of nats/Hz, where 1 bit $= \log_e(2)$ nats. 

\begin{theorem}[Uplink average rate]
\label{thm:uplinkAvgRate}
The average rate of a randomly chosen uplink user is given by $\bar{\tau}(\lambda,\alpha,\epsilon) = $ 
\begin{equation}
\int_0^\infty 2\pi\lambda r e^{-\pi\lambda r^2}\int_0^\infty e^{-\sigma^2  \mu \frac{e^{t}-1}{r^{\alpha(\epsilon-1)}}}\mathscr{L}_{I_z}\left(\mu \frac{e^{t}-1}{r^{\alpha(\epsilon-1)}}\right) \dd t\dd r,
\end{equation}
where $\mathscr{L}_{I_z}\left(s\right)$ is given by \eqref{eq:laplace2}.
\end{theorem}
\begin{IEEEproof}
Starting from the definition of $\bar{\tau}$ and $\sinr$, $\bar{\tau}(\lambda,\alpha,\epsilon)$
\begin{eqnarray}
\label{eq:rateeq}
&=& \mathbb{E}\left[\ln\left(1 + \sinr\right)\right] \nonumber \\
&=& \int_0^\infty \int_0^\infty \PP\left[\ln\left(1 + \sinr\right) > t\right]
\dd t f_R(r) \dd r\\
&=& \int_0^\infty f_R(r) \int_0^\infty \PP\left[\ln\left(1 + \frac{g \left(r^{\alpha(\epsilon-1)}\right)}{{\sigma}^2 + I_{\mathcal{Z}}}\right) > t\right]\dd t \dd r\nonumber\\
&=& \int_0^\infty f_R(r) \int_0^\infty \PP\left[ g> \frac{\left(e^{t}-1\right)({\sigma}^2 + I_{\mathcal{Z}})}{r^{\alpha(\epsilon-1)} }\right]\dd t \dd r\nonumber\\
&=&  \int_0^\infty f_R(r) \int_0^\infty e^{-s\sigma^2}\mathscr{L}_{I_z}\left(s\right) \dd t\dd r\nonumber,
\end{eqnarray}
where $f_R(r) = 2\pi\lambda r e^{-\pi\lambda r^2}$ and $s = \mu \frac{e^{t}-1}{r^{\alpha(\epsilon-1)}}$. The derivation of $\mathscr{L}_{I_z}\left(s\right)$ follows from Theorem \ref{thm:uplink1tier} and has either the form of \eqref{eq:laplacePPP} or \eqref{eq:laplaceUniform} depending on whether the random variable $R_z$ is used to model regular or irregular network.
\end{IEEEproof}

In Fig. \ref{fig:ulRate}, we plot the average user rate expressions using the Rayleigh assumption for $R_z$ as a function of $\epsilon$ for pathloss values of $\alpha = 2.5$, $3.25$, and $4$, transmit power $\mu^{-1} = 200$ mW, and no noise or $\sigma^2 = -104$dBm. The average rate increases with $\alpha$ over all values of $\epsilon$. Since the computed average rates are for a randomly chosen user anywhere in the network, the effects of power control on the high, medium, and low $\sinr$ users is combined into a single value. Thus as $\epsilon$ increases, the rate decreases due to the loss in rate for some users whose transmit power is reduced, which is not overcome on average by the reduction in interference and increased rate for other users, especially those near the cell-edge. We also note that for a dense deployment with $\lambda = .24$ base station/$\textrm{km}^2$, the no noise approximation is very tight.

\begin{figure} [t]
\begin{center}
   \includegraphics[width=\columnwidth]{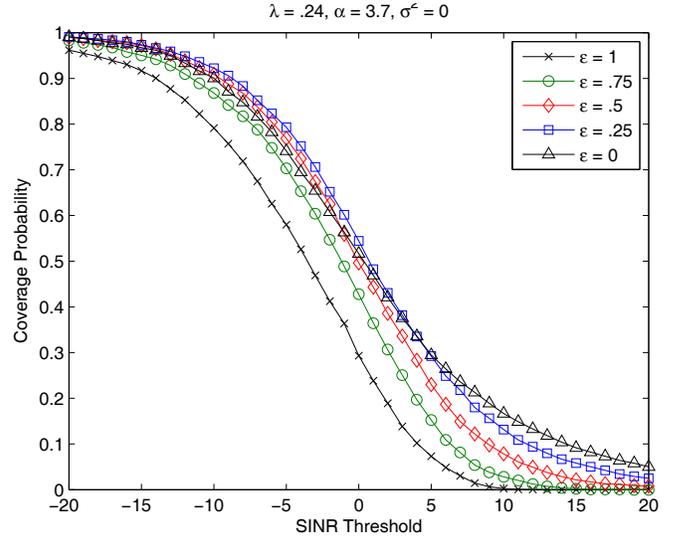}
   \caption{A plot of the uplink coverage probability for a PPP network with stochastic fractional power control and a range of $\epsilon$ values.}
   \label{fig:epsComp}
   \end{center}
\end{figure}

\begin{table}[ht] \caption{System Parameters}
\label{tab:LTEparams}
\begin{center}
{\normalsize
  \begin{tabular}{@{}p{0pt}@{}|l|l|}
    \hline 
    & Bandwidth &  10 MHz  \\
    \hline 
    & base station density &  .24 BS/$\textrm{km}^2$  \\
    \hline
    & User distribution &  uniform \\
    \hline
    & Pathloss (dB) &  $37\log(d)$, d = distance in meters  \\
    \hline
    & Downlink Tx Power & 45 dBm (30 W) \\
    \hline 
    & Uplink Max Tx Power & 23 dBm (200 mW)  \\  
    \hline 
    & FPC $\epsilon$ &  0.6, 0.8, 1.0  \\
    \hline 
    & Noise Power Density &  -174 dBm/Hz  \\
    \hline 
  \end{tabular}
}
\end{center} \end{table}

\subsection{Downlink vs. Uplink Coverage}
Another immediate application of the model is to consider the difference in coverage between the downlink and the uplink for the same network topology. We consider the randomly chosen user's $\sinr$ distribution in the downlink based on the model presented in \cite{AndrewsJ2011} for a network whose BSs are distributed according to a PPP with density $\lambda$. Before we make the comparison, it is important to highlight that the base station distribution in the proposed uplink model is not PPP, simply by the way it is constructed from the point process of the mobiles. Although an interesting problem in itself, the goal is not to characterize the base station point process and hence we numerically compare the downlink $\sinr$ distribution of the following two cases to justify the direct comparison of the uplink results proposed in this paper with the downlink results of~\cite{AndrewsJ2011}:
\begin{itemize}
\item Downlink: User-PPP: This corresponds to the case where we start with the base station locations that result from the proposed uplink location model, i.e., the mobiles form a PPP and each base station is located uniformly in the Voronoi cell of the mobile it is serving. The downlink $\sinr$ is then numerically evaluated at a randomly chosen mobile user assuming all the base stations transmit at the same power.
\item Downlink: BS-PPP: In this case, we consider the randomly chosen user's $\sinr$ distribution in the downlink based on the model presented in \cite{AndrewsJ2011} where the base station locations are modeled as a PPP with density $\lambda$. The coverage probability assuming all the base stations transmit at the same power is
\begin{equation}
p_c(T,\lambda,\alpha) = \pi \lambda \int_0^\infty  e^{-\pi \lambda v (1 + \rho(T,\alpha)) -\mu T \sigma^2 v^{\alpha/2}} \dd v,
\label{eq:dlPc}
\end{equation}
where
\begin{equation}
\rho(T,\alpha)= T^{2/\alpha}\int_{T^{-2/\alpha}}^\infty \frac{1}{1+u^{\alpha/2}} \dd u.
\label{eq:dlRho}
\end{equation}
\end{itemize}
In Fig. \ref{fig:DvU}, we plot the $\sinr$ distribution for both these cases and observe that the two models are close enough to facilitate the direct comparison of the downlink analytical results derived in \cite{AndrewsJ2011} and the uplink ones derived in this paper.

The downlink coverage probability, which assumes constant power transmissions across the network, has two major terms. The first term is based on $\rho(T,\alpha)$ and is not dependent on transmit power and represents the interference-limited contribution to the $\sinr$, while the second term depends on the noise power $\sigma^2$ and transmit power $\mu^{-1}$, representing the noise-limited part of the $\sinr$. However, the use of fractional power control in the uplink significantly changes the shape of the distribution of the interference power. This can be seen in both the noise-limited term with $\sigma^2$ and the Laplace transform of the interference, which are dependent on $\epsilon$ and the distribution of $R_z$ since the transmit powers of the mobiles throughout the network are not constant, but highly variable unlike the downlink.

Fig. \ref{fig:DvU} plots the two cases with the system parameters given in Table \ref{tab:LTEparams}, which are standard assumptions for a LTE-based cellular network \cite{Orange2011,3GPP2009,Ghosh2010}. We first note that the gap between the downlink results based on the BS-PPP and User-PPP assumptions is not very large which is reasonable given the prior evaluation of Assumption 2(a). Secondly we consider the disparity between the $\sinr$ distributions, especially for large $\sinr$ values. One reason for the uplink's lower coverage is due to the mismatch in transmit power compared to the downlink. Additionally at the high $\sinr$ values, the use of larger $\epsilon$ values also impacts the coverage probability since the users closest to their serving base stations greatly reduce their transmit power relative to the users at the cell-edge. The impact of $\epsilon$ is investigated in further detail in the following section.

The notion of disparate uplink and downlink coverage regions has fundamental consequences on the system design of cellular networks, different from those of wireless LANs, for example, which have much smaller coverage regions and typically do not have as significant hardware distinctions between the different network devices. For example, scenarios wherein the mobile user may be able to decode the downlink transmissions, but unable to connect via the uplink will impact handoff algorithms between base stations. One advantage of having a unified framework for uplink and downlink coverage is the ability to evaluate and optimize as function of the relevant system parameters. The aim may be to determine network and system parameters such that the uplink and  downlink are balanced or investigating tradeoffs between capacity and coverage enhancements at the base station and mobile terminal, respectively.  

\subsection{Fractional Power Control}
As mentioned previously, the primary motivations for fractional power control in the cellular uplink are to provide beneficial coverage improvements for the lowest-percentile users, who are typically at the cell-edge, and to manage average transmit power of battery-powered mobile devices. In practical cellular systems such as LTE, fractional power control parameters are network-specific and not user-specific, thus there needs to be some optimization performed to select parameters that can provide acceptable performance for the majority of users and provide a high overall system capacity \cite{Ericsson2007, Ericsson2008}. Fig. \ref{fig:epsComp} gives the coverage probability distributions as a function of the fractional power control factor $\epsilon$ for a network topology given by Table \ref{tab:LTEparams}. The baseline case of fixed transmit power for all users ($\epsilon = 0$) does not provide the lowest overall coverage probabilities, but does provide the greatest coverage probability for the highest $\sinr$ thresholds. The largest coverage probability for users in the lower 50 percentile is given by $\epsilon = 0.25$, followed by $\epsilon = 0.5$, giving gains over fixed transmit power before crossing below the $\epsilon = 0$ curve at 5 and 0 dB respectively. We also note that for the very low $\sinr$ thresholds $<-10$ dB, the difference in coverage probability for $\epsilon = 0, 0.25,$ and $0.5$ is negligible. As $\epsilon$ increases, the coverage probability curves shift lower with $\epsilon = .75$ providing much lower coverage probability than fixed transmit power, especially for $\sinr$ thresholds $> 5$ dB. Full pathloss inversion, $\epsilon = 1$ power control shows an even more significant reduction in coverage across all $\sinr$ thresholds. 

\begin{figure} [!ht]
\begin{center}
   \includegraphics[width=\columnwidth]{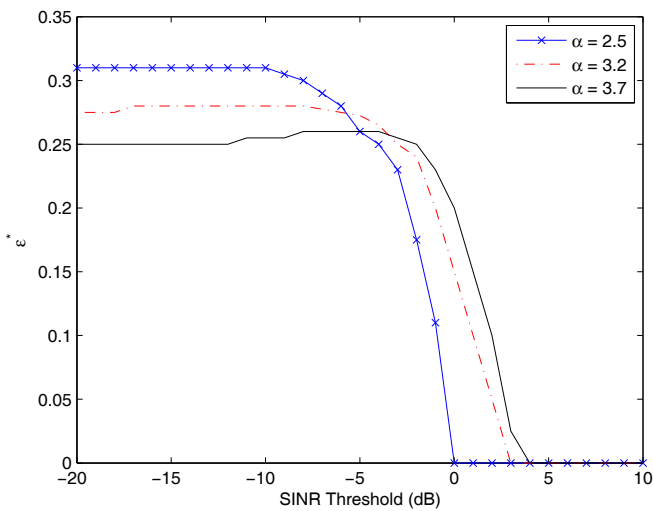}
   \caption{A plot of the coverage-maximizing $\hat{\epsilon}$ for a given $\sinr$ threshold value, $\lambda = .24$, and $\alpha = 2.5, 3.2$, and $3.7$.}
   \label{fig:epsOpt}
   \end{center}
\end{figure}

In Fig. \ref{fig:epsOpt}, we plot the value of $\epsilon$ that maximizes the coverage probability for a given $\sinr$ target $T$, denoted as $\hat{\epsilon}$ for $\alpha = 2.5, 3.2,$ and $3.7$. In other words, the value of $\hat{\epsilon}$ is determined according to $\hat{\epsilon}(T,\alpha,\lambda) = $
\begin{equation}
\label{eq:epsMax}
\arg\max_{\epsilon} 2\pi\lambda \int_0^\infty r e^{-\pi\lambda r^2-\mu Tr^{\alpha(1-\epsilon)}\sigma^2}\mathscr{L}_{I_z}\left(\mu Tr^{\alpha(1-\epsilon)}\right) \dd r,
\end{equation}
where the Laplace transform of the interference is given by \eqref{eq:laplace}.

This gives insight into the selection of $\epsilon$ from coverage probability maximization perspective. An interesting observation is that there are two distinct regions denoted in each plot by plateaus of near constant $\hat{\epsilon}$. For users with low $\sinr$ a moderate value of $\epsilon = .25$ to $.3$ provides the greatest gains while for users with high $\sinr$, the $\sinr$ is maximized by transmitting with the maximum power and $\epsilon = 0$. We note that there is some sensitivity to the pathloss exponent $\alpha$, in all three cases, the transition between the two regions is fairly steep in its slope, with an approximately 5 dB range, while as $\alpha$ increases the $\hat{\epsilon}$ for the low $\sinr$ region decreases while the $\sinr$ transition threshold between the two regions increases slightly. Additionally, this dual-regime behavior for fractional power control in uplink cellular networks differs from the behavior of power control in other classes of wireless networks, notably ad-hoc wireless networks, which were shown to have an optimal value of $\epsilon = .5$ \cite{AdHocPc2008,FlashLinq2010}. 

These observed effects of fractional power control in Fig. \ref{fig:epsComp} and Fig. \ref{fig:epsOpt} can be understood by focusing on the gains perceived by users close to their desired base station relative to those at the edge and their interdependency. Cell-interior users typically experience good RF conditions and are not as susceptible to interference as users at the cell edge. Instead, they are more noise-limited which means a reduction in their transmit power reduces their achievable $\snr$. This is especially true under pathloss-based power control, since high values of $\epsilon$ reserve the greatest transmit power for users with large pathloss. Thus transmitting at full power ($\epsilon = 0$) is the $\sinr$-optimizing strategy.

Cell-edge users, however, are more fundamentally interference limited, and the use of pathloss-proportional power control results in a decrease in the transmit power of interfering cell-center users and a relative increase in their transmit power (to overcome the larger cell-edge pathloss) benefiting their $\sinr$. This disparity becomes more pronounced with high $\epsilon$. As a result, there is a complex trade off in the reduction of interference from neighboring cell-center users and increased interference by mobiles at the cell edge, which gives rise to the observed intermediate range of $\epsilon$ values providing the highest gains for the majority of users. In effect, full-inversion power control performs a reordering of $\sinr$s between cell-edge and cell-interior users which does not provide system-wide gains. An advantage of the PPP uplink cellular model is that it captures the relevant system and topology parameters necessary for system designers to determine operating thresholds for a given range of parameters. 

\begin{figure} [!ht]
\begin{center}
   \includegraphics[width=\columnwidth]{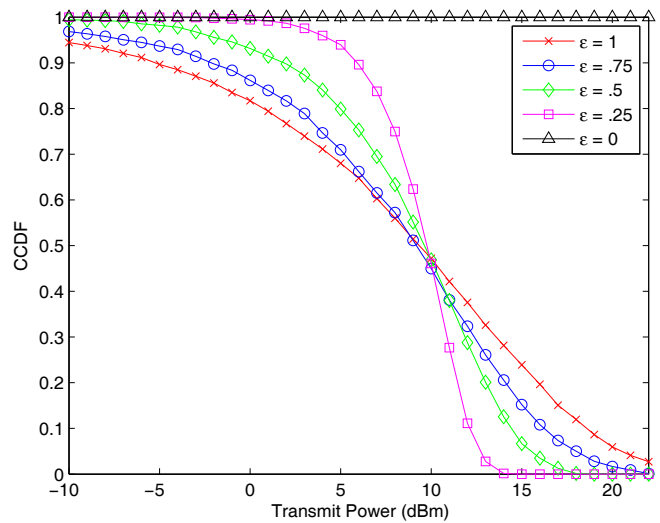}
   \caption{The CCDF of the average transmit power per mobile as a function of $\epsilon$ with $\lambda = .24$, $P_{max} = 23$dBm, $\mu^{-1} = 10$dBm, and $\alpha = 3.7$ pathloss factor.}
   \label{fig:txPwr}
   \end{center}
\end{figure}

Fig. \ref{fig:txPwr} gives the overall transmit power utilization of mobiles in the network as a function of $\epsilon$ with a maximum transmit power of $23$ dBm and an average transmit power of $\mu^{-1} = 10$ dBm. Clearly the transmit powers of the mobile users are greatly reduced with the introduction of power control. For high values of $\epsilon$ we note that 10-15\% of the users have transmit power less than $0$ dBm, which is a $23$ dB reduction in power compared to the maximum transmit power. For this reason, proposed system guidelines for the uplink may wish to choose $\epsilon$ to balance the metrics of coverage and battery utilization.

\section{Conclusion} 
This work has presented tractable expressions for the coverage probability and average rate in the cellular uplink, which are applicable both to uniform and irregular network topologies. The expressions are based on a novel analytical model utilizing the spatial Poisson process and are solely a function of the network topology and system design parameters including $\sinr$ targets, base station density, and fractional power control parameters. The presented results provide insight into the differences of downlink and uplink performance expectations and the tradeoff between using fractional power control to benefit cell-edge users and reducing overall power utilization by mobiles.

A major arena for future work is to understand how these dynamics are enhanced or differ for heterogeneous network topologies \cite{Dhillon2011,Gora2010}. The nested and overlapping nature of coverage regions for multiple tiers of access points is expected to have a large impact on the overall interference distribution and a range of hardware requirements and use cases may lead to unique system design tradeoffs not experienced in single-tier networks.

\section{Acknowledgments}

The authors thank F. Baccelli of The University of Texas at Austin and A. Ghosh of AT\&T Labs for their valuable inputs and insightful discussions.

\bibliographystyle{IEEEtran}
\bibliography{Paper-TW-Mar-12-0325.R2}

\begin{thebibliography}{10}
\providecommand{\url}[1]{#1}
\csname url@samestyle\endcsname
\providecommand{\newblock}{\relax}
\providecommand{\bibinfo}[2]{#2}
\providecommand{\BIBentrySTDinterwordspacing}{\spaceskip=0pt\relax}
\providecommand{\BIBentryALTinterwordstretchfactor}{4}
\providecommand{\BIBentryALTinterwordspacing}{\spaceskip=\fontdimen2\font plus
\BIBentryALTinterwordstretchfactor\fontdimen3\font minus
  \fontdimen4\font\relax}
\providecommand{\BIBforeignlanguage}[2]{{%
\expandafter\ifx\csname l@#1\endcsname\relax
\typeout{** WARNING: IEEEtran.bst: No hyphenation pattern has been}%
\typeout{** loaded for the language `#1'. Using the pattern for}%
\typeout{** the default language instead.}%
\else
\language=\csname l@#1\endcsname
\fi
#2}}
\providecommand{\BIBdecl}{\relax}
\BIBdecl

\bibitem{DhiNovC2012}
H.~S. Dhillon, T.~D. Novlan, and J.~G. Andrews, ``Coverage probability of
  uplink cellular networks,'' in \emph{Proc., IEEE Global Telecomm.
  Conference}, Anaheim, CA, 2012.

\bibitem{Wyner94}
A.~Wyner, ``Shannon-theoretic approach to a {Gaussian} cellular multiple-access
  channel,'' \emph{IEEE Transactions on Information Theory}, vol.~40, no.~6,
  pp. 1713--1727, Nov. 1994.

\bibitem{Wyner2000}
O.~Somekh and S.~Shamai, ``Shannon-theoretic approach to a {Gaussian} cellular
  multiple-access channel with fading,'' \emph{IEEE Transactions on Information
  Theory}, vol.~46, no.~4, pp. 1401--1425, July 2000.

\bibitem{Sin2006}
C.~Sin and V.~Lau, ``On the theoretical analysis of optimal cellular systems
  design with multi-user detection in slow flat fading channel - uplink
  analysis,'' in \emph{Proc. IEEE Vehicular Technology Conference}, vol.~1, May
  2006, pp. 182--186.

\bibitem{Simeone2009}
O.~Simeone, O.~Somekh, H.~Poor, and S.~Shamai, ``Local base station cooperation
  via finite-capacity links for the uplink of linear cellular networks,''
  \emph{IEEE Transactions on Information Theory}, vol.~55, no.~1, pp. 190--204,
  Jan. 2009.

\bibitem{Sanderovich2009}
A.~Sanderovich, O.~Somekh, H.~Poor, and S.~Shamai, ``Uplink macro diversity of
  limited backhaul cellular network,'' \emph{IEEE Transactions on Information
  Theory}, vol.~55, no.~8, pp. 3457--3478, Aug. 2009.

\bibitem{Levy2010}
N.~Levy and S.~Shamai, ``Information theoretic aspects of users' activity in a
  {Wyner}-like cellular model,'' \emph{IEEE Transactions on Information
  Theory}, vol.~56, no.~5, pp. 2241--2248, May 2010.

\bibitem{Onireti2011}
O.~Onireti, F.~Heliot, and M.~Imran, ``Closed-form approximation for the
  trade-off between energy efficiency and spectral efficiency in the uplink of
  cellular network,'' \emph{Proc. European Wireless Conference}, pp. 1 --6,
  Apr. 2011.

\bibitem{Jiaming2011}
J.~Xu, J.~Zhang, and J.~Andrews, ``On the accuracy of the {Wyner} model in
  cellular networks,'' \emph{IEEE Transactions on Wireless Communications},
  vol.~10, no.~9, pp. 3098--3109, Sep. 2011.

\bibitem{Dohler2006}
M.~Nawrocki, M.~Dohler, and A.~H. Aghvami, \emph{Understanding {UMTS} Radio
  Network Modelling, Planning and Automated Optimisation}.\hskip 1em plus 0.5em
  minus 0.4em\relax John Wiley \& Sons, Ltd, 2006.

\bibitem{Qualcomm2011}
\BIBentryALTinterwordspacing
Qualcomm, ``{LTE} advanced: heterogeneous networks,'' \emph{white paper}, Jan
  2011. [Online]. Available:
  \url{http://qualcomm.com/documents/files/lte-advanced-heterogeneous-networks.pdf}
\BIBentrySTDinterwordspacing

\bibitem{Haenggi2009}
M.~Haenggi, J.~Andrews, F.~Baccelli, O.~Dousse, and M.~Franceschetti,
  ``Stochastic geometry and random graphs for the analysis and design of
  wireless networks,'' \emph{IEEE Journal on Sel. Areas in Communications},
  vol.~27, no.~7, pp. 1029--1046, Sept. 2009.

\bibitem{Baccelli1995}
F.~Baccelli, M.~Klein, M.~Lebourges, and S.~Zuyev, ``Stochastic geometry and
  architecture of communication networks,'' \emph{J. Telecommunication
  Systems}, vol.~7, no.~1, pp. 209--227, 1997.

\bibitem{Brown2000}
T.~Brown, ``Cellular performance bounds via shotgun cellular systems,''
  \emph{IEEE Journal on Sel. Areas in Communications}, vol.~18, no.~11, pp.
  2443--2455, Nov. 2000.

\bibitem{AndrewsJ2011}
J.~Andrews, F.~Baccelli, and R.~Ganti, ``A tractable approach to coverage and
  rate in cellular networks,'' \emph{IEEE Transactions on Communications},
  vol.~59, no.~11, pp. 3122--3134, Nov. 2011.

\bibitem{MukherjeeWCDMA2011}
S.~Mukherjee, ``Analysis of {UE} outage probability and macrocellular traffic
  offloading for {WCDMA} macro network with femto overlay under closed and open
  access,'' in \emph{Proc., IEEE International Conference on Communications}.

\bibitem{MukherjeeLTE2011}
------, ``{UE} coverage in {LTE} macro network with mixed {CGS} and open access
  femto overlay,'' in \emph{Proc., IEEE International Conference on
  Communications}.

\bibitem{Dhillon2011}
H.~S. Dhillon, R.~K. Ganti, F.~Baccelli, and J.~G. Andrews, ``Modeling and
  analysis of {K}-tier downlink heterogeneous cellular networks,'' \emph{IEEE
  Journal on Selected Areas in Comm.}, Apr. 2012.

\bibitem{Novlan2011}
T.~Novlan, R.~Ganti, A.~Ghosh, and J.~Andrews, ``Analytical evaluation of
  fractional frequency reuse for {OFDMA} cellular networks,'' \emph{IEEE
  Transactions on Wireless Communications}, vol.~10, no.~12, pp. 4294--4305,
  Dec. 2011.

\bibitem{Novlan2012}
------, ``Analytical evaluation of fractional frequency reuse for heterogeneous
  cellular networks,'' \emph{IEEE Transactions on Communications}, vol.~60,
  no.~7, pp. 2029--2039, July 2012.

\bibitem{HetNetCommMag2012}
A.~Ghosh, N.~Mangalvedhe, R.~Ratasuk, B.~Mondal, M.~Cudak, E.~Visotsky,
  T.~Thomas, J.~Andrews, P.~Xia, H.~Jo, H.~Dhillon, and T.~Novlan,
  ``Heterogeneous cellular networks: From theory to practice,'' \emph{IEEE
  Communications Magazine}, vol.~50, no.~6, pp. 54 --64, june 2012.

\bibitem{TaylorC2012}
D.~Taylor, H.~S. Dhillon, T.~D. Novlan, and J.~G. Andrews, ``Pairwise
  interaction processes for modeling cellular network topology,'' in
  \emph{Proc., IEEE Global Telecomm. Conference}, Anaheim, CA, 2012.

\bibitem{Govindasamy2011}
S.~Govindasamy and D.~H. Staelin, ``Asymptotic spectral efficiency of the
  uplink in spatially distributed wireless networks with multi-antenna base
  stations,'' \emph{CoRR}, vol. abs/1102.1232, 2011.

\bibitem{Whitehead1993}
J.~Whitehead, ``Signal-level-based dynamic power control for co-channel
  interference management,'' in \emph{Proc. IEEE Vehicular Technology
  Conference}, may 1993, pp. 499 --502.

\bibitem{Yates1995}
R.~Yates, ``A framework for uplink power control in cellular radio systems,''
  \emph{IEEE Journal on Selected Areas in Communications}, vol.~13, no.~7, pp.
  1341--1347, Sep. 1995.

\bibitem{Holma2001}
H.~Holma and A.~Toskala, \emph{{WCDMA} for {UMTS}: Radio Access for Third
  Generation Mobile Communications, Revised Edition}.\hskip 1em plus 0.5em
  minus 0.4em\relax Wiley, 2001.

\bibitem{Herdtner2000}
J.~Herdtner and E.~Chong, ``Analysis of a class of distributed asynchronous
  power control algorithms for cellular wireless systems,'' \emph{IEEE Journal
  on Selected Areas in Communications}, vol.~18, no.~3, pp. 436--446, Mar.
  2000.

\bibitem{Kim2001}
D.~Kim, ``On the convergence of fixed-step power control algorithms with binary
  feedback for mobile communication systems,'' \emph{IEEE Transactions on
  Communications}, vol.~49, no.~2, pp. 249--252, Feb. 2001.

\bibitem{Agrawal2005}
A.~Agrawal, J.~Andrews, J.~Cioffi, and T.~Meng, ``Iterative power control for
  imperfect successive interference cancellation,'' \emph{IEEE Transactions on
  Wireless Communications}, vol.~4, no.~3, pp. 878--884, May 2005.

\bibitem{Ericsson2007}
Ericsson, ``R1-074850: Uplink power control for {E-UTRA} - range and
  representation of {P0},'' \emph{3{GPP} {TSG} {RAN} {WG}1 Meeting \#51},
  Novemeber 2007.

\bibitem{Ericsson2008}
A.~Simonsson and A.~Furuskar, ``Uplink power control in {LTE} - overview and
  performance, principles and benefits of utilizing rather than compensating
  for {SINR} variations,'' in \emph{Proc. IEEE Vehicular Technology
  Conference}, Sept. 2008, pp. 1 --5.

\bibitem{Mullner2009}
R.~Mullner, C.~Ball, K.~Ivanov, J.~Lienhart, and P.~Hric, ``Contrasting
  open-loop and closed-loop power control performance in {UTRAN} {LTE} uplink
  by {UE} trace analysis,'' in \emph{Proc. IEEE International Conference on
  Communications}, June 2009, pp. 1--6.

\bibitem{Castellanos2008}
C.~Castellanos, D.~Villa, C.~Rosa, K.~Pedersen, F.~Calabrese, P.~Michaelsen,
  and J.~Michel, ``Performance of uplink fractional power control in {UTRAN}
  {LTE},'' in \emph{Proc. IEEE Vehicular Technology Conference}, May 2008, pp.
  2517--2521.

\bibitem{Xiao2006}
W.~Xiao, R.~Ratasuk, A.~Ghosh, R.~Love, Y.~Sun, and R.~Nory, ``Uplink power
  control, interference coordination and resource allocation for {3GPP}
  {E-UTRA},'' in \emph{Proc. IEEE Vehicular Technology Conference}, Sept. 2006,
  pp. 1--5.

\bibitem{Rao2007}
A.~Rao, ``Reverse link power control for managing inter-cell interference in
  orthogonal multiple access systems,'' in \emph{Proc. IEEE Vehicular
  Technology Conference}, Oct. 2007, pp. 1837--1841.

\bibitem{Orange2011}
M.~Coupechoux and J.~Kelif, ``How to set the fractional power control
  compensation factor in {LTE}?'' in \emph{Proc. IEEE Sarnoff Symposium}, May
  2011, pp. 1--5.

\bibitem{Kelif2009}
J.~Kelif and M.~Coupechoux, ``Impact of topology and shadowing on the outage
  probability of cellular networks,'' in \emph{Proc. IEEE International
  Conference on Communications}, June 2009, pp. 1--6.

\bibitem{Stoyan1996}
D.~Stoyan, W.~Kendall, and J.~Mecke, \emph{Stochastic Geometry and Its
  Applications, 2nd Edition}.\hskip 1em plus 0.5em minus 0.4em\relax John Wiley
  and Sons, 1996.

\bibitem{3GPP2009}
3GPP, ``{TR} 36.814 v.1.5.2: Further advancements for {E-UTRA},'' Dec. 2009.

\bibitem{Tell99}
C.~Tellambura and A.~Annamalai, ``An unified numerical approach for computing
  the outage probability for mobile radio systems,'' \emph{IEEE Communications
  Letters}, vol.~3, no.~4, pp. 97--99, April 1999.

\bibitem{Berg04}
F.~Berggren and S.~Slimane, ``A simple bound on the outage probability with
  lognormally distributed interferers,'' \emph{IEEE Communications Letters},
  vol.~8, no.~5, pp. 271 -- 273, May 2004.

\bibitem{Ghosh2010}
A.~Ghosh, J.~Zhang, J.~G. Andrews, and R.~Muhamed, \emph{Fundamentals of
  {LTE}}.\hskip 1em plus 0.5em minus 0.4em\relax Prentice Hall, 2010.

\bibitem{AdHocPc2008}
N.~Jindal, S.~Weber, and J.~Andrews, ``Fractional power control for
  decentralized wireless networks,'' \emph{IEEE Transactions on Wireless
  Communications}, vol.~7, no.~12, pp. 5482 --5492, Dec. 2008.

\bibitem{FlashLinq2010}
X.~Wu, S.~Tavildar, S.~Shakkottai, T.~Richardson, J.~Li, R.~Laroia, and
  A.~Jovicic, ``{FlashLinQ}: A synchronous distributed scheduler for
  peer-to-peer ad hoc networks,'' in \emph{Allerton Conference on
  Communication, Control, and Computing}, Oct. 2010, pp. 514--521.

\bibitem{Gora2010}
J.~Gora, K.~Pedersen, A.~Szufarska, and S.~Strzyz, ``Cell-specific uplink power
  control for heterogeneous networks in {LTE},'' in \emph{Proc. IEEE Vehicular
  Technology Conference Fall}, Sept. 2010, pp. 1--5.

\end{thebibliography}

\end{document}